\documentclass{aa}
\usepackage{graphicx,epsfig}
\newcommand{\chem}[2]{$\rm{}^{#1}\kern-0.8pt#2$}
\newcommand{\chim}[2]{\rm{}^{#1}\kern-0.8pt#2}
\newcommand{\reac}[6]{$\rm\,{}^{#1}\kern-0.8pt{#2}\,({#3}\,,{#4})\,
           {}^{#5}\kern-0.8pt{#6}\,$}

\begin{document}


\title{The puzzle of the synthesis of the rare nuclide \chem{138}{La}}

\author{S. Goriely, M. Arnould, I. Borzov and M. Rayet} 
\institute{Institut d'Astronomie et d'Astrophysique, Universit\'e Libre de Bruxelles,
CP 226, B-1050 Brussels, Belgium}

\date{Received date; accepted date}

\abstract{
The calculations of the p-process in the O/Ne layers of Type II supernovae are quite
successful in reproducung the solar system content of p-nuclides. They predict, however, a
significant underproduction of the rare odd-odd nuclide \chem{138}{La}. A model for the
explosion of a 25 $M_{\odot}$ star with solar metallicity is used to suggest that
$\nu_e$-captures on \chem{138}{Ba} may well be its most efficient production mechanism.
The responsibility of an inadequate prediction of the \chem{138}{La} and
\chem{139}{La} photodisintegration rates in the too low production of \chem{138}{La} is
also examined quantitatively. A detailed discussion of the theoretical
uncertainties in these rates suggest that the required rate changes are probably
too high to be fully plausible. Their measurement would be most welcome. They would help
disentangling the relative contributions of thermonuclear and neutrino processes to the
\chem{138}{La} production.
}

\maketitle

\section{Introduction}

The odd-odd  neutron-deficient heavy nuclides \chem{138}{La} and isomeric  
\chem{180}{Ta}$^m$ are among the rarest solar system
species (no information exists for other locations), with \chem{138}{La}/\chem{139}{La}
$\approx 10^{-3}$ and \chem{180}{Ta}$^m$/\chem{181}{Ta} $\approx 10^{-4}$. In spite of their
scarcity, their origin has long been a puzzle. As
initially claimed by Prantzos et al. (1990) and confirmed by Rayet et al. (1995; hereafter
RAHPN), \chem{180}{Ta}$^m$ appears to be a natural product of the p-process in the
O/Ne-rich layers of Type II supernovae (SNII). In contrast, \chem{138}{La} is underproduced
in all p-process calculations performed so far (e.g. Fig.~1 of Arnould et al. 2001).

In view of the low \chem{138}{La} abundance, it has been attempted to explain its
production by non-thermonuclear processes involving either stellar energetic particles
(Audouze 1970) or neutral current neutrino-induced transmutations (Woosley et al. 1990).
The former mechanism is predicted not to be efficient enough, while the latter is found by
Woosley et al. (1990) to be able to overproduce the solar
\chem{138}{La}/\chem{139}{La} ratio by a factor of about 50. This prediction has to be
taken with some care, however, especially  in view of the qualitative nature of the
evaluation.

In a one-dimensional $Z = Z_\odot$ $M = 25$ M$_\odot$ SNII model,
\chem{138}{La} is predicted to be produced only at peak temperatures around $2.4 \pm
0.1 \times 10^9$~K (Arnould et al. 2001) from a subtle balance between its main
production by \reac{139}{La}{\gamma}{n}{138}{La} and its main destruction by
\reac{138}{La}{\gamma}{n}{137}{La}. The resulting abundances cannot account for the solar
system \chem{138}{La} amount. The same conclusion holds for all the stars in the $13
\leq M/$M$_\odot \leq 25$ examined by RAHPN. This situation might of course just 
result from inadequate astrophysics and/or nuclear physics inputs. On the
astrophysics side, one might incriminate an uncertain prediction of the evolution of
the thermodynamic conditions of the \chem{138}{La} producing layers during the explosion.
Modifying these conditions is unlikely, however, to cure the \chem{138}{La}
underproduction. For any astrophysically plausible conditions,  the SNII
layers releasing the highest \chem{138}{La} yields are also those
overproducing even more significantly heavier p-nuclides as \chem{156}{Dy},
\chem{162}{Er} or \chem{168}{Yb}. Of course, it remains to be seen if the situation could
be drastically modified if multi-dimensional effects were duly taken into account. On the
nuclear physics side, one has to be aware of the fact that the
\chem{138}{La} yield predictions rely entirely on theoretical nuclear reaction rates. One is
thus entitled to wonder about the sensitivity of the computed
\chem{138}{La} underproduction to the nuclear uncertainties that affect the production and
destruction channels. One of our aims is to provide the first quantitative examination of this
question (see also Arnould et al. 2001).

Our second aim is to analyse on more quantitative grounds than Woosley et al.
(1990) the possibility of producing \chem{138}{La} at a level compatible with the solar
system abundances through neutrino nucleosynthesis ($\nu$-process). To this end, the same $M
= 25$ M$_\odot$ SN model as considered above is selected. The RAHPN p-process network
is augmented with the neutrino and anti-neutrino charged-current inelastic scatterings
off the network nuclei, and with the neutrino neutral current inelastic scatterings off the
Ba, La and Ce isotopes.
 
Section 2 analyses the purely thermonuclear p-process production of \chem{138}{La}. A brief
description of the adopted astrophysical model and input nuclear physics is followed by the
evaluation of the impact on the calculated \chem{138}{La} yields of changes in the
\reac{139}{La}{\gamma}{n}{138}{La} and \reac{138}{La}{\gamma}{n}{137}{La} rates. We also
examine the plausibility of the rate modifications that are needed in order
to avoid an unsatisfactory \chem{138}{La} underproduction. The neutrino nucleosynthesis
contribution to the p-nuclides, and in particular to \chem{138}{La}, is discussed in
Sect.~3. Conclusions are drawn in Sect. 4.

\section{\chem{138}{La} and the p-process}

 
The p-process calculations reported here are based on a model for a $Z_{\odot}$ 8
M$_{\odot}$ helium star (main sequence mass of about 25 M$_{\odot}$) already considered by
RAHPN. It is  evolved from the beginning of core He burning to SN explosion (for details, see
Hashimoto 1995). We
consider here as P-Process Layers (PPLs) 25 O/Ne-rich zones with explosion temperatures
peaking in the  (1.7-3.3)$\times 10^9$ K range. Their total mass is approximately
0.75~M$_{\odot}$. The deepest PPL is located at a mass of about 1.94~M$_{\odot}$, which is
far enough from the mass cut for all the nuclides produced in this region to be ejected
during the explosion. 

The abundances of the s-process seeds for the production of the p-nuclides are calculated
with the NACRE `adopted' \reac{22}{Ne}{\alpha}{n}{25}{Mg} rate (Angulo et al. 1999), as
described by Costa et al. (2000; see the distribution labelled $R_1$ in their Fig.~2).

The p-process reaction network includes some 2500 stable and neutron-deficient nuclides
with $Z \le 84$. The n-, p- and $\alpha$-capture reactions on all nuclei are
considered, as well as their ($\gamma$,n), ($\gamma$,p) and ($\gamma,\alpha$)
photodisintegrations. The nuclear reaction rates are the NACRE `adopted' ones for
charged particle captures by nuclei up to \chem{28}{Si} (Angulo et al. 1999).  For
heavier targets, the rates predicted by the latest version of the Hauser-Feshbach code
MOST (Gor\-iely 2001) are adopted (the NACRE and MOST rates are available in
the Brussels Nuclear Astrophysics Library http://www-astro.ulb.ac.be). This version
benefits in particular from an improved description of the nuclear ground state
properties derived from the microscopic Hartree-Fock method (Goriely et al. 2001), as
well as from a more reliable nuclear level density prescription based on the
microscopic statistical model (Demetriou \& Goriely 2001). Finally, the experimentally-based
neutron capture rates of Bao et al. (2000) are used.
 
As in RAHPN, the computed abundance of the p-nuclide $i$ produced in the PPLs of the 
$Z_{\odot}$ 25 M$_\odot$ star considered here is represented by its mean overproduction
factor $\left< F_i \right> = \left< X_i \right> /X_{i,\odot}$, where $X_{i,\odot}$ is its
solar mass fraction (Anders \& Grevesse 1989), and
\begin{equation}
\left<X_i\right> = 
\frac{1}{M_p} \sum_{n \geq 1} (X_{i,n} + X_{i,n-1}) (M_n - M_{n-1})/2,
\end{equation}
where $X_{i,n}$ is the mass fraction of isotope $i$ at the mass coordinate $M_n$,
$M_p = \sum_{n\geq 1} (M_n - M_{n-1})$ is the total mass of the PPLs, the sum running
over all the PPLs ($n= 1$ corresponds to the bottom layer).  An overproduction factor
averaged over all 35 p-nuclei is calculated as $F_0 = \sum_i \left< F_i \right>/35$, and is
a measure of the global p-nuclide enrichment in the PPLs. 
So, if the computed  
abundances were exactly solar,  
$\left< F_i \right>/F_0$ would be equal to unity for all $i$.
 Fig.~1 indicates that most
of these factors lie in the 0.3 to 3 range. The cases
for which an underproduction is found are discussed by RAHPN or by Costa et al. (2000) for
\chem{92,94}{Mo} and \chem{96,98}{Ru}. Only the \chem{138}{La} problem is tackled here. 
 
\begin{figure}
\centerline{\epsfig{figure=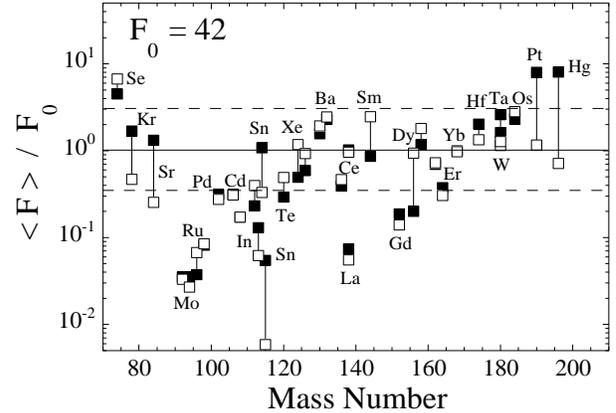,height=5.5cm,width=8cm}}
\caption{Normalised p-nuclide overproduction factors  $\left< F_i \right>/F_0$ ($F_0 = 42$)
obtained for the $Z = Z_\odot$ 25 M$_\odot$ model star with the nominal MOST rates
and in absence of  neutrino nucleosynthesis (open squares). For comparison, black squares
designate the factors calculated by RAHPN with the use of a different set of nuclear reaction
rates. The dotted horizontal lines delineate the $0.3 \leq \left< F_i \right>/F_0 \leq 3$
range.}
\end{figure}

 
We turn our special attention to the impact of alterations in the MOST rates for
\reac{137}{La}{n}{\gamma}{138}{La} and \reac{138}{La}{n}{\gamma}{139}{La}. The
Maxwellian-averaged cross sections at $T=2.5~10^9$~K are predicted to be 123 and
62~mbarn, respectively. These are used to calculate the
reverse
\chem{138}{La} and
\chem{139}{La} photodisintegrations of direct interest in the \chem{138}{La} production
by the application of the detailed balance theorem. More specifically, we examine the
extent to which the rate of
\reac{137}{La}{n}{\gamma}{138}{La} has to be {\it decreased} and the one of
\reac{138}{La}{n}{\gamma}{139}{La}  {\it increased} in order to bring the
\chem{138}{La} overproduction at levels comparable with those of the neighbouring
p-nuclides. In the following, the factors of decrease and increase will be noted
$F_{\rm down}$ and $F_{\rm up}$, respectively. The corresponding reverse
photodisintegration rates are decreased and increased by the same factors. In our numerical
tests, $F_{\rm down}$ and $F_{\rm up}$ are selected to vary independently in the 1 to 10
range.

Fig.~2 shows the ratio $R_{138} = \left< F(^{138}{\rm Ce})\right>/\left<
F(^{138}{\rm La})\right>$ obtained in this test. The choice of \chem{138}{Ce} as the
normalizing p-nuclide is dictated by the fact that it is produced in an amount close to
the average value $F_0$, so that $R_{138}$ gives a good representation of
the \chem{138}{La} production by the p-process. It is seen that $R_{138}$ is pushed
inside the $0.3 \leq \left< F \right>/F_0 \leq 3$ range represented in Fig.~1 only
for  
$F_{\rm down} \times F_{\rm up} \approx 20$ to 25. It remains to be seen if such changes
are physically plausible.

Many of the input data necessary to calculate reaction rates with MOST have been
measured for
\chem{139}{La}. This concerns in particular giant dipole energies and widths, or the
neutron resonance spacings. In contrast, very little is known experimentally for
\chem{137}{La} or \chem{138}{La}, so that their predicted neutron capture rates are likely
to be less reliable. In order to evaluate the rate uncertainties, a series of 
calculations have been performed with values of some basic input quantities differing
from their nominal values to an extent which is considered as reasonable (in view of the
trends and systematics for neighbouring nuclei). Of course, the compatibility with the
experimental constraints, if any, is always imposed.

This analysis clearly demonstrates that the main sources of uncertainties lie in the nuclear
level densities, $\gamma$-ray strength functions and neutron optical potentials. Related
errors in the capture rate predictions can amount to about a factor of 2 for
\chem{138}{La}(n,$\gamma$)\chem{139}{La} at the temperature of relevance (2.4~$10^9$~K) for
the \chem{138}{La} synthesis. In view of scarcer experimental information, larger
uncertainties affect the \chem{137}{La}(n,$\gamma$)\chem{138}{La} rate. Even so, it is
unlikely that the product $F_{\rm down} \times F_{\rm up}$ introduced above can
realistically reach the requested value of about 20 to 25 in order for the underproduction
of \chem {138}{La} with respect to \chem{138}{Ce} not to exceed a factor of 3 (see Fig.~2).
Of course, only experimental determinations of the
\chem{138}{La} and \chem{139}{La} photodisintegration rates could really put this
theoretical conclusion on a safe footing. Such experiments are feasible today (Vogt et al.
2001) and are strongly encouraged.

\begin{figure}
\centerline{\epsfig{figure=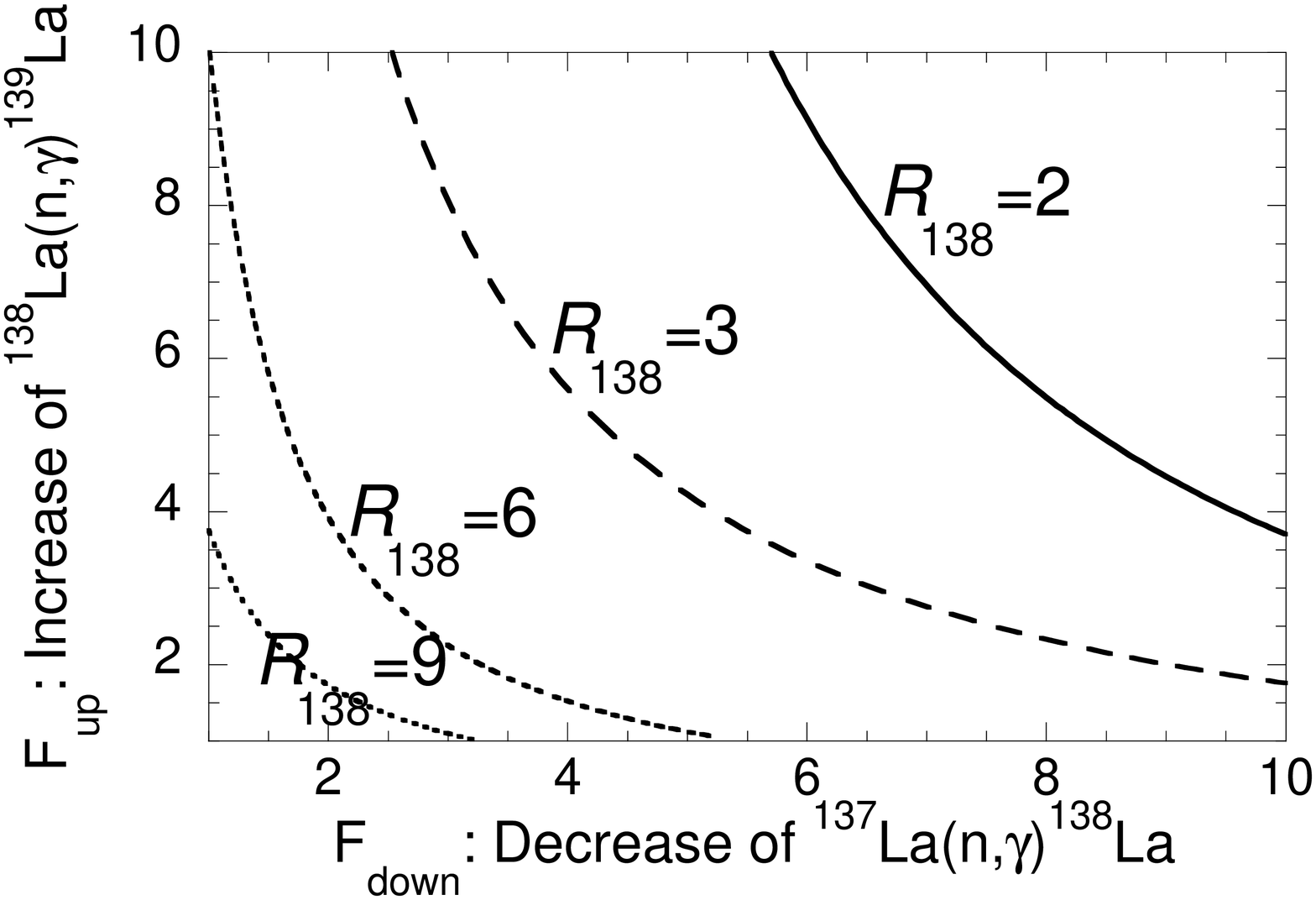,height=6cm,width=9cm}}
\caption{\chem{138}{Ce} to \chem{138}{La} overproduction ratio, $R_{138}$, for
different values of $F_{\rm down}$ and $F_{\rm up}$.}
 \label{F2}
\end{figure}

\section{The production of \chem{138}{La} by neutrino captures}

The calculations of Sect.~2 neglect the irradiation of the PPLs by the neutrinos
streaming out of the proto-neutron star. In order to examine the impact of the neutrino
interactions on the p-nuclide yields, and in particular on \chem{138}{La}, the reaction
network used in Sect.~2 is augmented with all neutrino-induced reactions (charged and
neutral currents) up to Kr (all rates are from Hoffman \& Woosley 1992), as
well as all charged-current neutrino and antineutrino scatterings up to Po. The
neutral-current scatterings off nuclei heavier than Kr are considered only for the Ba, La
and Ce isotopes. 

The neutrino and antineutrino scattering cross sections are averaged over supernova
(anti)neutrino spectra that are approximated by zero-degeneracy Fermi-Dirac distributions
corresponding to typical temperatures $T_{\nu} = 8, 5$ and 4 MeV for the $\nu_x$ (where
$x$ stands for $\mu$ and $\tau$ (anti)neutrinos), $\bar \nu_e$ and $\nu_e$, respectively
(Fuller \& Meyer 1995). The allowed transitions for the charged-current (anti)neutrino
scatterings are treated fully microscopically within the cQRPA approximation (Borzov et al.
1995, Borzov \& Goriely 2000) and on grounds of the ETFSI ground state description  (Aboussir
et al. 1995). The Gamow-Teller (GT) strength function accounts for the transitions to the
daughter nucleus ground and low-lying states, to the GT resonance (GTR) and to the particle
continuum region above the GTR. The entire ($N-Z$) Fermi strength is contained in the
isobaric analog resonance (IAR), its energy being taken from the experimental systematics of
Coulomb displacement energies. It has to be noted that the IAR  and GTR usually appear as
decay states for $N < Z$ nuclei. In such conditions, $\bar \nu$-captures dominate
$\nu$-captures (the situation is the reverse in $N > Z$ nuclei, where the IAR
and the main part of the GT strength are excited via $\nu$-capture).  

As the average $\nu_e (\bar \nu_e)$ energies are low ($E < 16$ MeV), allowed transitions
are expected to dominate the scattering processes. For nuclei in the neighbourhood of
\chem{138}{La}, we just take into account the contribution from the forbidden
transitions in a very rough way through the increase by 25\% of the microscopically
calculated $\nu(\bar \nu)$-capture cross sections for allowed transitions (Hektor et al.
2000). In particular, the resulting averaged $\nu_e$-capture cross section by
\chem{138}{Ba} is estimated to be 7.5~$10^{-41}$~cm$^2$ at $T_{\nu_e}=4$~MeV. The neutral
current $\nu$-scattering contribution is also estimated in a rough way by assuming an
approximate scaling of the ($\nu_x,\nu{_x} \prime$) cross section with mass number. At
$T_{\nu_x}=8$~MeV, the excitation cross section followed by neutron emission of the La
isotopes is $1.2~10^{-41}$~cm$^2$.

The energy-integrated number flux of neutrinos of type $\nu$ at a radius $r_7$ (expressed in
units of 100~km) is estimated in terms of the neutrino-sphere radius and temperature $T_\nu$
and of the $\nu$-neutrino luminosity $L_\nu$ to be (Fuller \& Meyer 1995) 

\begin{equation}
\Phi_{\nu}=1.58 \times 10^{41}~{\rm cm^{-2}~s^{-1}} (\frac{L_{\nu}}{10^{51}{\rm
ergs~s^{-1}}})~(\frac{{\rm MeV}}{kT_{\nu}}) ~\frac{1}{r_7^2}.
\end{equation}

\noindent In this expression, $L_{\nu}$ is assumed to vary between about
$10^{51}$ and $10^{52}~{\rm ergs~s^{-1}}$. This quite broad range of values is selected in
order to accomodate to the many intricacies and uncertainties in the detailed neutrino
fluxes emanating during the whole nucleosynthesis episode from the nascent neutron star
underlying the O/Ne-rich zones of interest here. Note that $\Phi_\nu$ [Eq.~(2)] depends on
$r_7$, which varies with time in all the considered layers. 
Let us also emphasize that no oscillation between neutrino species is taken into account. Would
$\nu_e$'s be converted to heavier species by matter-induced oscillations, larger $\nu(\bar
\nu)$-capture cross sections on bare nuclei could be expected (mainly due to the 
contribution of forbidden transitions).\footnote{This work was completed by the time the SNO
collaboration has announced that their neutrino data were supporting the existence of
neutrino oscillations (Giles 2001). A quantitative analysis of the impact of this exciting
discovery on the $\nu$-synthesis of \chem{138}{La} has clearly to be postponed}

\begin{figure}
\centerline{\epsfig{figure=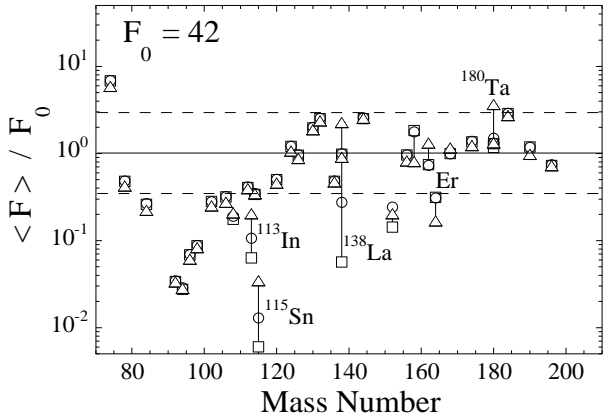,height=5.5cm,width=8cm}}
\caption{Comparison of the p-process abundances obtained with and without
(squares) neutrino interactions (with standard MOST rates). Two sets of luminosities
$L_\nu[10^{51}~{\rm ergs~s^{-1}}]=$(3, 4, 16) (circles) and (30, 40, 160) for  ($\nu_e, \bar
\nu_e$,
$\nu_x$)(triangles) are adopted for the calculations with neutrinos. }
 \label{F3}
\end{figure}

Fig.~\ref{F3} shows the impact of the neutrino interactions on the p-nuclide production for
the two sets of typical luminosities (Fuller \& Meyer 1995) $L_\nu[10^{51}~{\rm
ergs~s^{-1}}]=$ (3, 4, 16) and (30, 40, 160) for ($\nu_e,
\bar \nu_e$ and $\nu_x$). These two $L_\nu$ combinations lead to increases of the
\chem{138}{La} production with respect to the case without neutrinos by factors of 4.8
and 36, respectively. This enhanced synthesis originates entirely from the
$\nu_e$-capture by
\chem{138}{Ba}, the neutral current scatterings on \chem{139}{La} being found to have a
negligible impact. Despite relatively similar cross sections, the \chem{138}{Ba}
$\nu_e$-capture is found particularly efficient due to the large initial
\chem{138}{Ba} abundance (about 10 times the \chem{139}{La} abundance). For the high
luminosity set, the
$\nu_e$-captures also enhance the production of
\chem{113}{In},
\chem{115}{Sn},
\chem{162}{Er} and
\chem{180}{Ta}. Despite the numerous uncertainties still affecting supernova models and
the neutrino physics in supernovae (spectra, luminosity, temperature, oscillation,
interaction cross sections,
\dots), $\nu_e$-captures appear so far to be the most efficient production mechanism of the
solar \chem{138}{La}. 

\section{Conclusions}

Woosley et al. (1990) have considered as an `intriguing possibility' the production of
\chem{138}{La} by (charged current) $\nu_e$-captures on \chem{138}{Ba}. We
provide the first quantitative support to this possibility through the use of a
`realistic' one-dimensional SNII model and of an extended network of nuclear
and neutrino reactions the rates of which benefit from a series of
improvements. Still, astrophysics and nuclear/neutrino physics uncertainties remain and
forbid this conclusion to be as strong as one might want.

The responsibility of an inadequate prediction of the \reac{138}{La}{\gamma}{n}{137}{La}
and \reac{139}{La}{\gamma}{n}{138}{La} photodisintegration rates in the too low production of
\chem{138}{La} is also examined quantitatively. Clearly, a suitable \chem{138}{La}
production could be obtained by adequate changes in the nominal rates. However, it is
concluded from a detailed study of the theoretical uncertainties in these rates that the
level of the required changes is probably too high to be fully plausible. An experimental
study of the above-mentioned key rates is no longer out of reach and is eagerly awaited.
These measurements would certainly help greatly disentangling the relative thermonuclear
and neutrino synthesis contributions to one of the rarest nuclides in nature.

\acknowledgements{S.G. and M.R. are FNRS research associates }

\end{document}